\PassOptionsToPackage{merge}{nab} 
\documentclass{webofc}
\usepackage[varg]{txfonts}   
\usepackage{placeins}
\usepackage{cleveref}
\newcommand{\dd}{\textrm{d}}
\newcommand{\ZE}{Z_\textrm{E}}
\newcommand{\Tr}{\textrm{Tr}\,}
\newcommand{\tauf}{\tau_F}

\newcommand{\als}{\alpha_{\text{s}}}
\def\lsim{\mathrel{\raise.3ex\hbox{$<$\kern-.75em\lower1ex\hbox{$\sim$}}}}
\def\gsim{\mathrel{\raise.3ex\hbox{$>$\kern-.75em\lower1ex\hbox{$\sim$}}}}

\begin{document}
\title{The static force from generalized Wilson loops \\using gradient flow}

\author{\firstname{Viljami} \lastname{Leino}\inst{1}\fnsep\thanks{\email{viljami.leino@tum.de}} \and
        \firstname{Nora} \lastname{Brambilla}\inst{1,2,3}\fnsep\thanks{\email{nora.brambilla@ph.tum.de}} \and
        \firstname{Julian} \lastname{Mayer-Steudte}\inst{1}\fnsep\thanks{\email{julian.mayer-steudte@tum.de}} \and
        \firstname{Antonio} \lastname{Vairo}\inst{1}\fnsep\thanks{\email{antonio.vairo@ph.tum.de}} 
}

\institute{Physik Department, Technische Universit\"at M\"unchen,\\
James-Franck-Strasse 1, 85748 Garching, Germany
\and
Institute for Advanced Study, Technische Universit\"at M\"unchen,\\
Lichtenbergstrasse 2 a, 85748 Garching, Germany
\and
Munich Data Science Institute, Technische Universit\"at M\"unchen, \\
Walther-von-Dyck-Strasse 10, 85748 Garching, Germany
          }

\abstract{%
  We explore a novel approach to compute the force between a static quark-antiquark pair with
  the gradient flow algorithm on the lattice. The approach is based on inserting a chromoelectric field in
  a Wilson loop. The renormalization issues, associated with the finite size of the chromoelectric field on
  the lattice, can be solved with the use of gradient flow. We compare numerical results for the
  flowed static potential to our previous measurement of the same observable without a gradient flow.
}
\maketitle
\section{Introduction}
\label{intro}
The static potential $V(r)$ between a static quark and an antiquark is a quantity that has been 
studied for a long time in QCD. The perturbative expansion for $V(r)$ is known up to $\mathrm{N}^3\mathrm{LO}$ accuracy.
The perturbative expression of the static potential at short distances 
allows one to accurately extract the strong coupling, $\als$, from lattice QCD computations of the same quantity.
This $\alpha_\mathrm{s}$ extraction is competitive with other existing determination from different observables~\cite{Aoki:2019cca}.

In lattice regularization, the static potential comes with a linear divergence proportional to the inverse 
lattice spacing $1/a$. This divergence is associated with contributions coming from the self-energy of the 
static quarks. The self-energy vanishes in dimensional regularization. 
In dimensional regularization, the perturbative expression for the static potential 
is, however, affected by a renormalon of order $\Lambda_\text{QCD}$~\cite{Pineda:1998id,Hoang:1998nz}.
Both the renormalon and the linear divergence on the lattice
can be absorbed into an additive constant. This constant contribution has no physical significance
and can be removed by considering the static force $F(r)=\partial_r V(r)$ instead.
The static force encodes the shape of $V(r)$ and carries all the physical information needed to extract $\als$, while being finite and renormalon free.

A precise determination of the static force from the static potential on the lattice can be challenging.
The errors associated with the numerical derivative of the data are directly proportional to the 
density of the data points. In quenched QCD simulations, the data can be dense enough for a reliable
definition of the numerical derivative~\cite{Necco:2001xg,Necco:2001gh}. 
In full QCD simulations, the data at short distances is, however,
still too sparse~\cite{Bazavov:2014soa}. 
This problem can be avoided by considering a recently suggested method~\cite{Vairo:2015vgb,Vairo:2016pxb},
where the force between a static quark and a static antiquark is measured directly on the lattice 
by considering a Wilson loop with a chromoelectric field $E$ inserted into one of the temporal Wilson lines.
A proof of concept study of this method has recently been carried out in~\cite{Brambilla:2019zqc,Brambilla:2021wqs}.

In the previous study~\cite{Brambilla:2019zqc,Brambilla:2021wqs}, 
it was found that the insertion of the chromoelectric field induces additional
self-energy contributions on the lattice. These self-energy contributions cause a slow convergence
towards the continuum limit. Such behavior is expected, as it is well known that
operators that involve components of the field strength tensor often come with sizable 
discretization errors associated with the slow convergence of lattice perturbation theory when expanded
with respect to the bare coupling. This issue can be reduced by introducing a multiplicative renormalization factor $Z_E$. 
This renormalization factor can be estimated in multiple different ways, 
see e.g, Refs.~\cite{Huntley:1986de,Lepage:1992xa, Bali:1997am,Koma:2006fw,Guazzini:2007bu,Christensen:2016wdo}.
In Ref~\cite{Brambilla:2021wqs}, we suggested a method to measure this renormalization factor non-perturbatively 
by considering the ratio of the static force defined from the Wilson loop with a chromoelectric field insertion 
and the static force defined from the finite difference of the static potential measured on the lattice.

An interesting alternative approach to this renormalization issue is offered by an algorithm known as the gradient flow~\cite{Luscher:2010iy}. 
In this proceeding, we study the static force measured on the lattice with the gradient flow 
and show that the gradient flow reduces sizably discretization effects  
related to insertion of components of the field strength tensor.
The original talk at the conference also summarized the results from Ref.~\cite{Brambilla:2021wqs}.
However, due to page limitations, this summary has been split to a separate proceeding contribution~\cite{Leino:2021bpz}.

The proceeding is organized as follows. In Section~\ref{sec:force} we introduce our observables.
The gradient flow algorithm is explained and applied to our observables in section~\ref{sec:flow}.
We discuss the simulation details and present the numerical results in section~\ref{sec:res}.

\section{Definition of the static force}\label{sec:force}
\label{forcesec}
The static potential between a static quark and an antiquark is related to the trace of the Wilson loop with temporal extent $T$ and spatial extent $r$,
\begin{equation}
\label{EQN764} {\rm Tr}\{{\rm P} \, W_{r \times T}\} = \Tr \bigg\{{\rm P} \, \exp\bigg(i g \oint_{r \times T} dx_\mu \, A_\mu(x)\bigg)\bigg\} \,,
\end{equation}
from where the static potential can be extracted on the lattice with the following limit
\begin{equation}
\label{EQN_V_lat} V(r,a) = \lim_{T \rightarrow \infty}  V_\text{eff}(r,T,a) \;, \quad V_\text{eff}(r,T,a) 
= -\frac{1}{a} \ln \frac{\langle \Tr \{{\rm P} \, W_{r \times (T+a)}\}\rangle}{\langle \Tr\{{\rm P} \, W_{r \times T}\}\rangle} \,.
\end{equation}
The traditional definition of the static force $F(r)$ via a finite difference is then,
\begin{equation}\label{eq:fdv}
F_{\partial V} (r,a) 
= \frac{V(r+a,a) - V(r-a,a)}{2 a} \,.
\end{equation}

An alternative method for computing the static force was proposed in Refs.~\cite{Vairo:2015vgb,Vairo:2016pxb},
\begin{equation}
\label{EQN_F} F(r) = 
\lim_{T \rightarrow \infty} -i \frac{\langle \Tr\{{\rm P} \, 
W_{r \times T} \,\hat{\mathbf{r}} \cdot g \mathbf{E}(\mathbf{r},t^\ast)\}\rangle}{\langle \Tr\{{\rm P} \, W_{r \times T}\} \rangle} \,.
\end{equation}
Here the chromoelectric field $\mathbf{E}(\mathbf{r},t^\ast)$ is inserted in one of the temporal Wilson lines at a fixed time $t^\ast$. 
The choice of $t^\ast$ is arbitrary, because the $t^\ast$ dependence of $\langle {\rm Tr}\{{\rm P} \, W_{r \times T} \, g E_j({\bf r},t^*)\} \rangle$  
disappears in the limit $T \rightarrow \infty$ if $t^\ast$ is kept constant.
In practice, for the purposes of lattice simulations, we set $t^\ast$ to be in the middle of the temporal Wilson line 
in order to maximize the distance from the temporal boundaries of the Wilson loop.

The chromoelectric field components are defined in terms of the field strength tensor as $E_j(x) = F_{j 0}(x)$.
These components have to be discretized on the lattice. We define the chromoelectric field using a so called cloverleaf formulation~\cite{Bali:1997am}:
\begin{equation}
g E_j = \frac{\Pi_{j 0} - \Pi^\dagger_{j 0}}{2 i a^2} + \mathcal{O}(a^2) \;, \quad
\label{EQN_cloverleaf} \Pi_{j 0} = \frac{P_{j,0} + P_{0,-j} + P_{-j,-0} + P_{-0,j}}{4}\,,
\end{equation}
where $P_{\mu,\nu} = 1 + i a^2 g F_{\mu \nu} = U_\mu(x) U_\nu(x+\hat{\mu}) U^\dagger_\mu(x+\hat{\nu}) U_\nu^\dagger(x)$ is the plaquette
defined as a product of link variables $U_\mu(x) = e^{i a g A_\mu(x)}$.
We then define an effective force to extract the static force from lattice simulations,
\begin{equation}
\label{eq:f1} F_E(r,a) = \lim_{T \rightarrow \infty} F_{E,\text{eff}}(r,T,a) \quad , \quad F_{E,\text{eff}}(r,T,a) 
                       = -i \frac{\langle \Tr\{{\rm P} \, W_{r \times T} \, \hat{\mathbf{r}} \cdot g \mathbf{E}(\mathbf{r},t^*)\}\rangle}{\langle \Tr\{{\rm P} \, W_{r\times T}\}\rangle}\,.
\end{equation}

\section{Gradient flow}\label{sec:flow}
The Yang-Mills gradient flow~\cite{Luscher:2010iy} is a smearing algorithm 
that evolves a given gauge field configuration towards the classical solution,
\begin{align}
\partial_{\tauf} B_{\tauf,\mu} &= -g^2 \frac{\delta S_\mathrm{YM}}{\delta B_{\tauf,\mu}} = D_{\tauf,\nu} G_{\tauf,\nu\mu} \label{eq:gff} \\
G_{\tauf,\mu\nu} &= \partial_{\mu} B_{\tauf,\nu} - \partial_{\nu} B_{\tauf,\mu} + \left[B_{\tauf,\mu},B_{\tauf,\nu}\right] \\
B_{0,\mu} &= A_{\mu}\,,
\end{align}
where $\tauf$ is the flow time and $S_\mathrm{YM}$ is the gauge action. 
For this study, we choose $S_\mathrm{YM}$ to be the Lüscher-Weisz action. 
On the lattice, the gauge fields are replaced with the link variables $U_{\mu}$ and the flow equation can be expressed as,
\begin{equation}\label{eq:dgfd}
\partial_{\tauf} U_{\mu}(\vec{x},\tauf) = -g_0^2 \left(\frac{\partial S_\mathrm{YM}[U]}{\partial U_{\mu}(\vec{x},\tauf)}\right) U_{\mu}(\vec{x},\tauf)\,.
\end{equation}
This equation smears the gauge links with a flow radius $\sqrt{8\tauf}$.

In perturbation theory, the flowed fields $B$ can be written in terms of the gluon fields $A$ by iteratively solving~\eqref{eq:gff}. 
This leads to a set of Feynman rules for the flowed fields. 
At the leading order of perturbation theory, the flow modifies the gluon propagator\,,
\begin{equation}
D(p,\tauf) = D(p,\tauf=0)e^{-2\tauf p^2} = \frac{1}{p^2}e^{-2\tauf p^2}\,.
\end{equation}
This allows a tree-level definition of the static potential and the static force at a finite flow time,
\begin{align}
V_\mathrm{continuum}^\mathrm{LO}(r,\tauf) &= -\frac{\als C_\mathrm{F}}{r} \mathrm{erf}(\frac{r}{\sqrt{8\tauf}}) \label{eq:C_V_f} \\
F_\mathrm{continuum}^\mathrm{LO}(r,\tauf) &= \frac{\als C_\mathrm{F}}{r^2} \left[\mathrm{erf}(\frac{r}{\sqrt{8\tauf}}) - \frac{r}{\sqrt{2\pi \tauf}} e^{-\frac{r^2}{8\tauf}}\right]\,,\label{eq:C_F_f}
\end{align}
where erf is the error function. 
For the purposes of this proceeding, the leading order result is enough.
However, we note that the static potential and the static force at finite flow time have been calculated up to one-loop level in Ref.~\cite{Brambilla:2021egm}.
This calculation will be important for the final extraction of the static force at finite flow time on the lattice.

On the lattice, we use different actions for the simulation itself (Wilson action) and the gradient flow (Lüscher-Weisz action).
In this case, the flowed propagator becomes~\cite{Fodor:2014cpa},
\begin{align}
D_\mathrm{total,lattice}(p,\tauf) &= e^{-\tauf D^{-1}_\mathrm{flow, lattice}(p,\tauf=0)} D_\mathrm{simulation, lattice}(p,\tauf=0) e^{-\tauf D^{-1}_\mathrm{flow, lattice}(p,\tauf=0)}\\
D_\mathrm{*,lattice}^{-1} &= 4\sum_{i=1}^{4} \left( \sin^2 \frac{p_i}{2} + c_w \sin^4 \frac{p_i}{2}\right)\,,
\end{align}
where $c_w=0$ for the Wilson action and $c_w=1/3$ for the Lüscher-Weisz action.
Furthermore, the static potential at a finite flow time is then given by
\begin{equation}\label{eq:L_V_f}
V_\mathrm{lattice}^\mathrm{LO}(r,\tauf) = -C_\mathrm{F} g^2 \int \frac{\dd^3 p}{(2\pi^3)} D_\mathrm{total,lattice}(p,\tauf)\,.
\end{equation}
The static force in the leading order of lattice perturbation theory is then given by a finite difference as defined in  Eq.~\eqref{eq:fdv}~\cite{Brambilla:2021wqs}.

Having defined the finite flow time static force in both continuum and lattice perturbation theories, we can improve the lattice results
for the static force at tree-level by defining an improved force,
\begin{equation}\label{eq_treelvlimp}
F^\mathrm{imp}_\mathrm{lattice}(r, \tauf, a) = \frac{F^\mathrm{LO}_\mathrm{continuum}(r,\tauf)}{F^\mathrm{LO}_\mathrm{lattice}(r,\tauf)} F_\mathrm{E}(r,\tauf, a)\,.
\end{equation}
Similarly, the static potential can be tree-level improved by taking a ratio of Eqs.~\eqref{eq:C_V_f} and~\eqref{eq:L_V_f}.

\section{Simulations and results}\label{sec:res}
We discretize the SU(3) Yang-Mills gauge theory using the standard Wilson plaquette action. Furthermore, for the flow equation we use the Lüscher-Weisz discretization.
We generate the gauge configurations using the heatbath and overrelaxation algorithms. 
On these configurations, we then integrate the discretized gradient flow~\eqref{eq:dgfd} using an optimized adaptive algorithm~\cite{Bazavov:2021pik}. 
We generate three ensembles with parameters and statistics listed in table~\ref{tab-1}, where the lattice spacing $a$ in physical units is related to the
gauge coupling with parameterization from Ref.~\cite{Necco:2001xg}. We have chosen the set of ensembles to be close to the set used in~\cite{Brambilla:2021wqs} to ease the
comparison with existing results.
\begin{table}
\centering
\begin{tabular}{cccc}
\hline
$\beta$ & $N_\sigma\times N_t$ & a[fm] & \# configurations  \\\hline
6.284 & $20\times 40$ & 0.060 & 1949  \\
6.481 & $26\times 56$ & 0.046 & 1999  \\
6.594 & $30\times 60$ & 0.040 & 1997  \\
\hline
\end{tabular}%
\caption{Gauge link ensembles and their parameters}
\label{tab-1} 
\end{table}

First, we test the renormalization property of the gradient flow. 
It is known that observables involving components of the field strength tensor often 
exhibit sizable discretization errors at the values of the gauge coupling typically used in numerical simulations.
This is due to a slow convergence of lattice perturbation theory, when expanded in the bare coupling~\cite{Lepage:1992xa}. 
The discretization errors can be reduced with a suitable multiplicative renormalization factor as discussed in Refs.~\cite{Huntley:1986de,Bali:1997am,Koma:2006fw,Guazzini:2007bu,Christensen:2016wdo}.
This renormalization factor is the main difference in discretization effects between the two definitions of the static force $F_E(r,a)$ and $F_{\partial V}(r,a)$ given by Eqs.~\eqref{eq:f1} and~\eqref{eq:fdv}
respectively. Hence, we can define a multiplicative improvement factor $Z_E$ that determines this renormalization,
\begin{equation}
\label{EQN_ZE} \ZE(a) = \frac{F_{\partial V}(r^\ast,a)}{F_E(r^\ast,a)} ,
\end{equation} 
where $r^\ast$ is an arbitrary separation. 
We use the tree-level improved~\eqref{eq_treelvlimp} static force and potential when taking the ratio. 

\begin{figure}[t]
\centering
\includegraphics[width=1.0\textwidth]{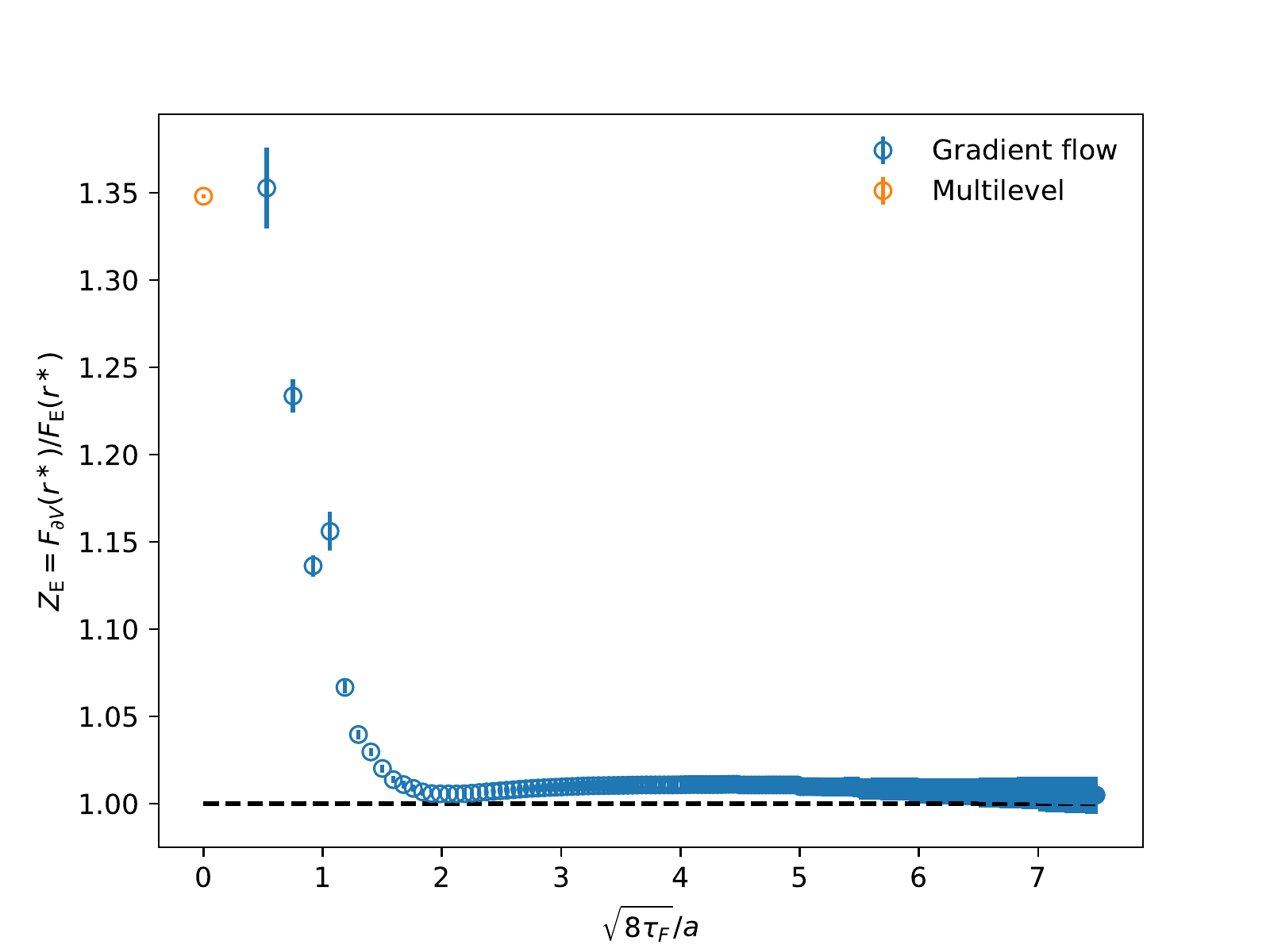}
\caption{The renormalization factor $Z_E = F_{\partial V}/F_E$ as a function of the flow radius $\sqrt{8\tauf}$.
The point in orange shows the measured $Z_E$ at zero flow time from Ref.~\cite{Brambilla:2021wqs}.}
\label{fig-1}  
\end{figure}

For each flow time $\tauf$, we vary $r^\ast$ over all values of $r$ and find a range of intermediate $r$ values, where the data can be described with a constant fit.
This constant then defines the renormalization factor $Z_E$ as a function of flow time.
The flow time dependent renormalization constant is shown in Fig.~\ref{fig-1}.
We observe that at zero flow time we replicate the result from the previous study~\cite{Brambilla:2021wqs}. 
As the flow time is increased, $Z_E$ decreases rapidly until settling to a constant value $Z_E\approx 1$ for $\sqrt{8 \tauf} > 1.6a$.
The renormalization factor becoming unity indicates that the gradient flow, indeed, removes the sizable discretization effects caused by the finite discretization of the chromoelectric field.
The remaining structure in $Z_E$ for $\sqrt{8 \tauf} > 1.6a$ is most likely due to underestimated systematic errors.

Next, we compare the gradient flow result to an existing zero flow time measurement of the derivative of the static potential. 
In Ref.~\cite{Brambilla:2021wqs}, we measured the static potential traditionally at zero flow time and performed a fit to the Cornell ansatz. 
The Cornell potential can then be differentiated analytically. We use this Cornell fit as our ground truth when testing the new static force formulations.

\begin{figure}[t]
\centering
\includegraphics[width=1.0\textwidth]{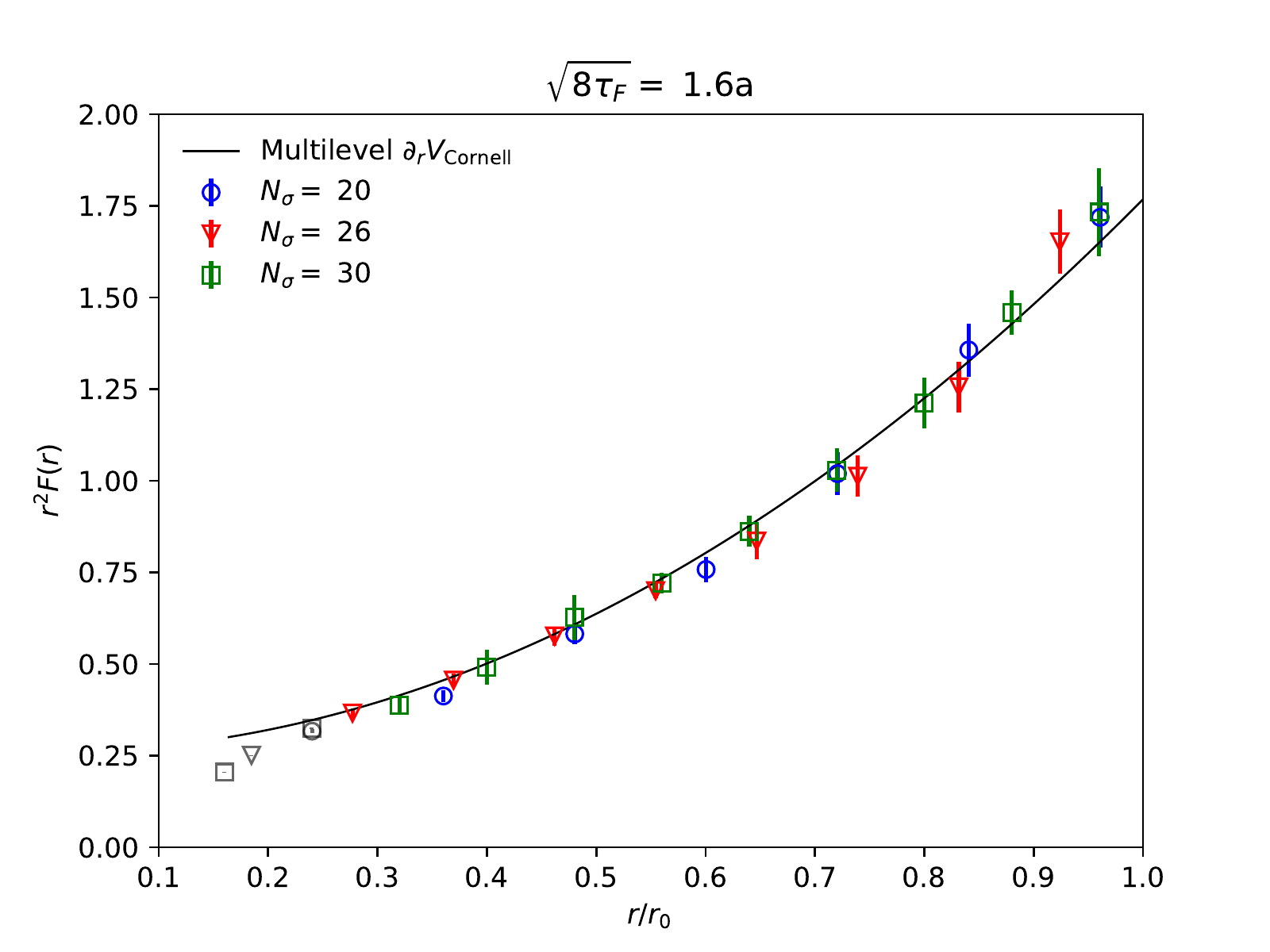}
\caption{The static force $r^2 F_E$ at flow time $\sqrt{8 \tauf} > 1.6a$ compared to 
$\partial_r V_\mathrm{Cronell}(r)$ from Ref.~\cite{Brambilla:2021wqs}.
The gray points present the data where the flow radius is much larger than the separation $r$.
}
\label{fig-2}  
\end{figure}

Ideally, one should perform the zero flow time limit only after the continuum limit has been taken. 
Unfortunately, we currently lack the required range of lattice spacings to confidently perform the continuum limit.
Hence, we cannot perform the zero flow time limit either.
Instead, we compare the zero flow time result to the static force measured at the smallest flow time $\sqrt{8 \tauf} > 1.6a$ where $Z_E \sim 1$.
This is shown in Fig.~\ref{fig-2}.
We see, that despite the lack of continuum and zero flow time limits, 
the static force from the gradient flow simulations agrees with the previous result for the 
derivative of the static potential measured at zero flow time.
Since the flow smears the gauge link, we expect to see enhanced discretization effects when the
flow radius grows to order of $\sqrt{8 \tauf}\gsim r/2$. These points are shown in gray in Fig.~\ref{fig-2},
where we observe indeed that these points slightly deviate from the expected behavior.

In summary, we have shown preliminary results of the static force measured with gradient flow from 
the expectation value of a Wilson loop with a chromoelectric field insertion. 
We observe that at finite flow times the renormalization constant $Z_E$, 
which removes the sizable discretization errors associated with the chromoelectric field insertion, becomes consistent with unity.
This shows in practice, the well known renormalization property of the gradient flow algorithm.
We have also shown an early comparison between our flowed data and the existing result at zero flow time.
The continuum and zero flow time limits are left for a later publication. 
These limits and the strong coupling extraction could be improved further with the recent perturbative calculation 
of the flowed static force at one loop level~\cite{Brambilla:2021egm}.

\FloatBarrier
\section*{\acknowledgementname}
This work has been supported by 
the Deutsche Forschungsgemeinschaft (DFG, German Research Foundation) cluster of excellence ORIGINS
that is funded by the DFG under Germany’s Excellence Strategy EXC-2094-390783311. 
The simulations were performed using the publicly available MILC code~\cite{MILC} and were carried
out on the computing facilities of the Computational Center for Particle and Astrophysics (C2PAP) 
of the cluster of excellence ORIGINS.

\bibliography{proceedings.bib}

\end{document}